\documentclass[journal]{IEEEtran}
\usepackage{cite}

\usepackage{subfigure}
\usepackage{graphicx}
\usepackage{graphics}
\usepackage{amssymb}
\usepackage{color}
\usepackage{booktabs}
\usepackage{threeparttable}

\usepackage[colorlinks=true, citecolor=blue]{hyperref}

\usepackage{amsthm}
\newtheorem{definition}{Definition}

\usepackage{changepage}

\usepackage{makecell}

\usepackage{tabularx}

\usepackage[switch,left]{lineno}
\usepackage{xcolor}

\usepackage{soul}

\ifCLASSINFOpdf
\else
\fi
\usepackage[cmex10]{amsmath}
\usepackage{array}
\usepackage{url}

\begin{document}

\twocolumn

\title{Transforming Police-Car Swerving for Mitigating Isolated Stop-and-Go Traffic Waves:
A Practice-Oriented Jam-Absorption Driving Strategy}

%
%
%

\author{Zhengbing~He, {\it Senior Member, IEEE}
\thanks{
Z. He is with the Faculty of Science and Engineering, University of Nottingham Ningbo China ({\it he.zb@hotmail.com})
}


}


%
%

{}
%



\maketitle

\begin{abstract}
Stop-and-go traffic waves, a major form of freeway congestion, impose severe and persistent adverse impacts, including reduced traffic efficiency, increased safety risks, and elevated vehicle emissions.
Among various freeway traffic management strategies, jam-absorption driving (JAD), in which a dedicated vehicle performs ``slow-in" and ``fast-out" maneuvers before being captured by a stop-and-go wave, has been proposed as a promising approach to suppressing the propagation of such waves.
However, most existing JAD strategies remain impractical, primarily due to the lack of consideration of implementation vehicles and operational conditions.
Inspired by real-world observations of police-car swerving behavior, this paper first introduces the Single-Vehicle Double-Detector Jam-Absorption Driving (SD-JAD) problem and then proposes a practical JAD strategy based on a definition of the JAD Triangle, transforming such behavior into a traffic control strategy capable of suppressing the propagation of an isolated stop-and-go wave.
Five key parameters that significantly affect the proposed strategy, namely JAD speed, inflow traffic speed, wave width, wave speed, and in-wave speed, are identified and systematically analyzed.
Using a SUMO-based simulation as an illustrative example, we further demonstrate how these parameters can be measured in practice using only two stationary roadside traffic detectors.
The results show that the proposed JAD strategy successfully suppresses the propagation of a stop-and-go wave without triggering secondary waves.
This paper is expected to take a significant step toward the practical implementation of JAD, advancing it from a theoretical concept to a feasible and deployable traffic management strategy.
To promote reproducibility in the transportation domain, we have also open-sourced all the code on our GitHub repository: \url{https://github.com/gotrafficgo/jam_absorption_driving}.

\end{abstract}

\begin{IEEEkeywords}
Traffic congestion, traffic control, travel time, vehicle dynamics, mobility, connected and automated vehicle

\end{IEEEkeywords}

\section{Introduction}\label{sec:Intro}

\IEEEPARstart{S}{top-and-go} traffic waves (also referred to as wide moving jams or traffic oscillations) are a common form of freeway congestion.
Such traffic patterns force vehicles to repeatedly accelerate and decelerate, resulting in reduced traffic efficiency, longer travel times \cite{Yildirimoglu2013,Zhang2017c}, increased fuel consumption and emissions \cite{Li2020,Jiang2025}, and a higher risk of traffic accidents \cite{Wang2009,Wang2024}.
Even worse, under sufficiently high traffic demand, stop-and-go waves can propagate and sustain themselves over tens of kilometers, further amplifying their adverse impacts.


Among freeway traffic management strategies, jam-absorption driving (JAD), in which a dedicated vehicle performs ``slow-in'' and ``fast-out'' maneuvers before being captured by a stop-and-go wave, has been proposed as a promising approach for preventing the propagation of such waves \cite{Beaty1998,Beaty2013}.
Early studies focused primarily on fundamental theoretical questions: Can JAD suppress stop-and-go waves? Will it trigger secondary waves? Initial analyses based on Kinematic Wave Theory and the Fundamental Diagram suggested that JAD can effectively suppress wave propagation \cite{JAD2013}. Microscopic traffic flow models, such as the Helly car-following model and the Intelligent Driver Model, were subsequently employed to capture vehicle dynamics more explicitly \cite{JAD2015,JAD2020b}, although their conclusions relied on relatively strong modeling assumptions.
Since 2017, research has gradually shifted toward more practical and implementation-oriented JAD strategies \cite{JAD2017}. Key issues include determining when to activate and deactivate a JAD vehicle and identifying the most suitable vehicle for intervention. To address these challenges, existing studies have incorporated wave propagation prediction \cite{JAD2017}, influential subspace concepts \cite{Jerath2015}, capacity-drop considerations \cite{JAD2021b}, multiple-wave suppression \cite{JAD2024b}, and traffic efficiency and safety optimization \cite{JAD2020,JAD2021,JAD2021b}. Simulations are typically used to demonstrate the effectiveness of JAD in improving traffic flow and suppressing stop-and-go waves.
For a more comprehensive review of recent developments in JAD, one may refer to our latest review article \cite{He2025}.

Generally speaking, although JAD strategies have been extensively studied, most assume the availability of Connected and Automated Vehicles (CAVs) to perform JAD tasks and full access to traffic information to determine a specific JAD plan. 
Unfortunately, even with the rapid development of CAV technology, fully controlling one or a few CAVs in real-world traffic to implement traffic management remains highly challenging and is {\bf rarely observed.}
Moreover, the issue of JAD triggering secondary waves has been raised in nearly all existing studies, yet few have actually addressed it in a systematic way. 
Therefore, while JAD is theoretically appealing, its practical implementation is still far from being realized.

\begin{figure*}[htbp]
    \centering
    \includegraphics[width=\linewidth]{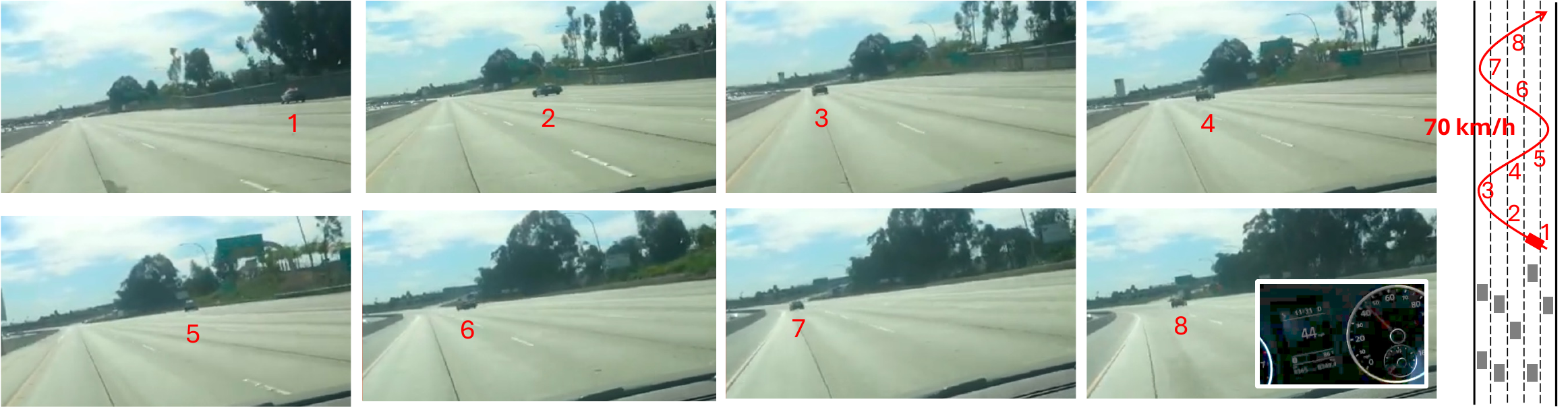}
    \caption{Police-car swerving at high speed on a freeway in California, USA (snapshots from the footage in \cite{Carpenter2025video}). 
    The footage shows that the police car was swerving across lanes at a speed of approximately 70 km/h (44 mph) for roughly one minute. Eventually, it caused the upstream traffic to merge into a queue at a relatively low speed, suggesting a possible objective of slowing down traffic inflow.}
    \label{fig:swerving}
\end{figure*}

Occasionally, we observed a type of police-car swerving behavior frequently appearing on social media. 
As shown in Figure~\ref{fig:swerving} and the footage in \cite{Carpenter2025video}, a police car was recorded maneuvering at high speed across all lanes on a freeway in California, USA, to control the following traffic. 
The speed may reach up to 70 km/h, and the duration is approximately 1 minute. 
Although the objective of the police car remains unclear to the authors and is likely aimed at slowing down traffic inflow, this real-world behavior demonstrates the practical feasibility of deploying a moving vehicle to regulate the speed of upstream traffic.

Building on this real-world observation, this paper aims to develop a practically feasible strategy for suppressing stop-and-go waves. 
To this end, we formulate a {\it Single-Vehicle Double-Detector Jam-Absorption Driving (SD-JAD)} problem, in which only two stationary roadside traffic detectors (upstream and downstream) and a single JAD vehicle are assumed to be available. 
Under these {\bf minimal conditions}, the JAD objective is to block the propagation of a stop-and-go wave, specifically an isolated wave, which is distinct from periodic stop-and-go waves associated with fixed bottlenecks.

We further propose a simple, first-order JAD strategy with a definition of the JAD Triangle\footnote{One of the anonymous reviewers comments as follows: {\it The ``JAD Triangle" concept provides an elegant, first-order geometric framework for determining deactivation points without resorting to overly complex second-order models that often fail to align with real-world physics.}} and identify five key parameters that significantly influence its performance, namely, {\it JAD speed}, {\it inflow traffic speed}, {\it wave width}, {\it wave speed}, and {\it in-wave speed}. 
The effects of these parameters on the strategy are thoroughly analyzed. 
Using SUMO-based simulations, we discuss how these parameters can be measured in practice with only two stationary traffic detectors, and demonstrate the effectiveness of the proposed JAD strategy.
The triggering of secondary waves is successfully prevented by taking traffic stability into account when selecting the JAD speed.
By leveraging police-car swerving behavior to implement JAD, we believe this work substantially advances JAD from a purely theoretical concept toward practical application.
Moreover, to promote reproducibility in the transportation domain, we have open-sourced all the code on our GitHub repository: \url{https://github.com/gotrafficgo/jam_absorption_driving}.

The remainder of the paper is organized as follows. 
Section~\ref{sec:Problem} introduces the SD-JAD problem. 
Section~\ref{sec:Question} proposes the JAD strategy and addresses several key questions. 
Section~\ref{sec:Analysis} presents a detailed theoretical analysis of the key parameters and validates the results through numerical experiments. 
Section~\ref{sec:Simulation} uses SUMO-based simulations to illustrate how these parameters can be measured in practice using only two stationary roadside traffic detectors and to further validate the effectiveness of the proposed strategy. 
Finally, conclusions are drawn in Section~\ref{sec:Conclusion}.

\section{Problem Statement}\label{sec:Problem}

Inspired by the real-world observation of police-car swerving on a freeway, as shown in Figure~\ref{fig:swerving} and in the footage \cite{Carpenter2025video}, we introduce the SD-JAD problem as follows.

\begin{adjustwidth}{6mm}{6mm}
\begin{definition}
\label{def:sd-jad}
{\bf Single-Vehicle Double-Detector Jam-Absorption Driving (SD-JAD)} \vspace{1mm}\\
It is a freeway traffic control problem in which only a single dedicated vehicle is employed to suppress the propagation of an incoming isolated stop-and-go wave.
The dedicated vehicle merges into the traffic stream from a fixed location (e.g., freeway shoulder or on-ramp), and both the dispatch decision and the detailed driving plan are determined solely based on information from a pair of detectors located upstream and downstream.
\end{definition}
\end{adjustwidth}
\vspace{3mm}
Note that the targeted wave is an {\bf isolated stop-and-go wave}, characterized by high-density free-flow conditions both upstream and downstream, such as the so-called phantom traffic jam. 
Such waves may originate from disturbances under high traffic demand. 
Many existing variable speed limit (VSL) strategies, such as the well-known SPECIALIST \cite{VSL2008,VSL2010}, are designed to address such types of waves \cite{He2025}. 
In contrast, the proposed strategy does not aim to resolve {\bf periodic stop-and-go waves} associated with fixed-location bottlenecks. 
However, many existing JAD studies do not explicitly clarify this distinction.

\begin{figure}[htbp]
    \centering
    \includegraphics[width=\linewidth]{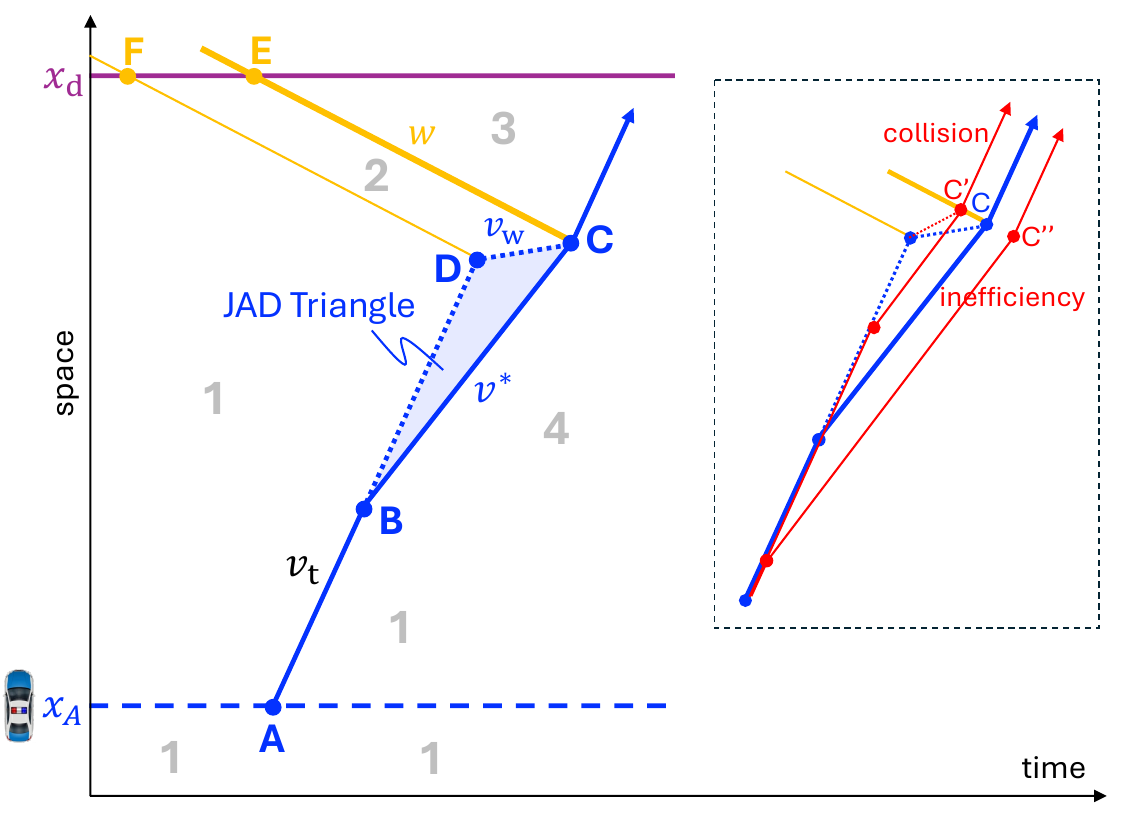}
    \caption{A schematic diagram of the Single-Vehicle Double-Detector Jam-Absorption Driving (SD-JAD) problem. Note that the gray numbers indicate traffic states.}
    \label{fig:problem}
\end{figure}

Figure~\ref{fig:problem} describes the problem, in which $x_A$ denotes the location of the standby dedicated vehicle for JAD; $x_{\mathrm d}$ and $x_{\mathrm u}$ denote the locations of the downstream and upstream detectors, respectively; $w$ denotes the stop-and-go wave speed; and $v_{\mathrm t}$ and $v^*$ denote the car-following speed and the pre-determined JAD speed of the dedicated vehicle, respectively.

There are a total of six critical time-space points:
\vspace{1mm}
\begin{adjustwidth}{2em}{0pt}
\begin{itemize}
	\item $A (t_A, x_A)$ is the point where the JAD vehicle enters the traffic stream;
	\item $B (t_B, x_B)$ is the point where the JAD vehicle starts to implement the JAD task with speed $v^*$;
	\item $C (t_C, x_C)$ is the point where the JAD vehicle stops implementing the JAD task, i.e., where the stop-and-go wave is terminated;
	\item $D (t_D, x_D)$ is the point where the JAD vehicle (or a regular vehicle) would reach the front of the stop-and-go wave without JAD intervention;
	\item $E (t_E, x_E)$ is the point where the temporal rear of the stop-and-go wave is detected by the downstream detector at location $x_{\mathrm d}$;
	\item $F (t_F, x_F)$ is the point where the temporal front of the stop-and-go wave is detected by the downstream detector at location $x_{\mathrm d}$.\vspace{1mm}		
\end{itemize}
\end{adjustwidth}
The JAD plan can be described as follows:
\begin{equation}\label{equ:JAD}
A\xrightarrow{v_{\mathrm t}} B \xrightarrow{v^*} C,
\end{equation}
where $A\xrightarrow{v_{\mathrm t}} B$ represents a process of floating within traffic and waiting for an appropriate moment, and $B \xrightarrow{v^*} C$ is the actual JAD operation.

We also define the triangle $BCD$ as the \textbf{JAD Triangle}, which directly determines the details of the JAD strategy.
The common condition for the JAD Triangle is
\begin{equation}\label{equ:common}
v_{\mathrm t} > v^* > v_{\mathrm w} \ge 0.
\end{equation}
Points $E$ and $F$ are associated with the stop-and-go wave detected by the downstream detector at $x_{\mathrm d}$, and thus $x_{\mathrm d} = x_E = x_F$.
The temporal width of the stop-and-go wave is denoted by $\Delta_{\mathrm w} = t_E - t_F$.

The key questions for solving the SD-JAD problem are as follows.
\vspace{1mm}
\begin{adjustwidth}{2em}{0pt}
\begin{enumerate}
	\item[Q1.] What is the JAD speed, i.e., $v_A$?
	\item[Q2.] What is the JAD plan, i.e., $A\xrightarrow{v_{\mathrm t}} B \xrightarrow{v^*} C$? \vspace{1mm}
\end{enumerate}
\end{adjustwidth}
It is worth noting that terminating the JAD task either earlier or later than point $C$ (e.g., at $C'$ or $C''$) may lead to safety risks or inefficiency, as shown in Fig.~\ref{fig:problem}.

The key assumptions are as follows.
\vspace{1mm}
\begin{adjustwidth}{2em}{0pt}
\begin{enumerate}
	\item[A1.] Stop-and-go waves propagate backward at an approximately constant speed, which is consistent with real-world observations \cite{Mauch2002a,Laval2010,Zheng2011a,Jiang2014,He2015a}.
	
	\item[A2.] The length of the backward-moving queue is also approximately constant, rather than continuously increasing, as shown in many real-world observations \cite{He2025}. If the queue shrinks (e.g., due to reduced inflow), there is no need to dispatch a JAD vehicle. This assumption can be relaxed by incorporating additional information to estimate variations in wave width.
	

	\item[A3.] For simplicity, we assume an instantaneous speed change in the theoretical analysis. In simulations or practice, a gradually changing speed would not significantly affect the results, as shown in \cite{JAD2017}.
\end{enumerate}
\end{adjustwidth}
\vspace{1mm}

\section{Key Questions and Solutions}\label{sec:Question}

\subsection{Q1: What is the JAD speed?}\label{sec:jad_speed}

First, the JAD speed should be lower than the speed of the surrounding or incoming traffic, i.e., $v^* < v_{\mathrm t}$, consistent with the original ``slow-in'' and ``fast-out'' concept of JAD.

Second, as extensively discussed in \cite{He2025}, a main potential hazard of JAD is the triggering of secondary waves. Therefore, avoiding the generation of secondary waves is a key consideration in determining the JAD speed.

\begin{figure}[htbp]
    \centering
    \includegraphics[width=0.7\linewidth]{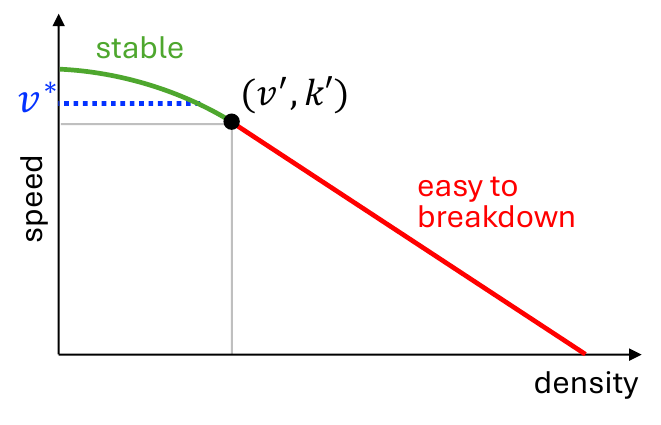}
    \caption{Speed-density relationship and traffic instability. $v'$ and $k'$ denote the critical traffic state separating stable and unstable traffic regimes.}
    \label{fig:jad_speed}
\end{figure}

According to fundamental traffic flow theory, secondary waves or traffic breakdowns are primarily associated with traffic instability. 
Generally speaking, low-density traffic is stable, while high-density traffic is prone to breakdown\footnote{
The notion of stability can differ depending on the perspective. From a microscopic or dynamical viewpoint, a high-density flow is linearly unstable, i.e., small perturbations can grow into stop-and-go waves. However, from a macroscopic or phase-based perspective, once traffic has entered a congested state, the flow pattern becomes relatively stable and does not further break down. In this study, the instability considered corresponds to the first case, i.e., from a microscopic viewpoint.
} (Figure~\ref{fig:jad_speed}).
There may exist a critical density (denoted by $k'$) and an associated critical speed (denoted by $v'$) that separate these states\footnote{The critical density and speed here may differ from those commonly used in traffic flow theory to distinguish free-flow and congested conditions.}.
Therefore, to avoid triggering secondary waves, the JAD speed should be higher than the critical speed, i.e., $v^* \ge v'$.

Unfortunately, determining the value of $v'$ in real-world traffic remains challenging, even though it is well understood that traffic instability is directly related to imperfections in human driving behavior and can be explained using traffic stability theory. 
The main difficulty arises from traffic heterogeneity, including both vehicle and driver behavior heterogeneity, which is highly complex and dynamic, making accurate estimation of $v'$ in real traffic extremely challenging. 
In this paper, as will be shown later, we determine this value through simulation. In practice, one could select a conservative value for $v^*$, i.e., a value sufficiently higher than the assumed $v'$ based on experience, which would greatly simplify the problem. 
This issue cannot be fully addressed within a single paper, and therefore more in-depth investigation of the estimation of $v'$ is expected.

\subsection{Q2: What is the JAD plan? Part I: Points $B$ and $C$}

As shown in Figure~\ref{fig:problem}, the JAD plan can be described by \eqref{equ:JAD}. We assume that the values of the following variables have been pre-determined:
\begin{itemize}
\item The pre-defined JAD speed $v^*$;
\item The surrounding or approaching traffic speed $v_{\mathrm t}$;
\item The temporal rear and front of a stop-and-go wave at a fixed location, detected by the downstream detector at $x_{\mathrm d}$, i.e., points $E$ and $F$;
\item The wave speed $w$.
\end{itemize}

The selection of point $A$ will be carefully discussed later in Section \ref{sec:Q2}.
Here, we simply assume that point $A$ is known, and then we only need to measure points $B$ and $C$.
Referring to Figure \ref{fig:problem}, from a geometric perspective, given point $F$ and slope $w$, we have line $DF$ expressed as follows:
\begin{equation}\label{equ:DF}
	DF: \   x = x_F + w (t - t_F).
\end{equation}
Given point $E$ and slope $w$, we have line $CE$ expressed as follows:
\begin{equation}\label{equ:CE}
	CE: \  x = x_E + w (t - t_E).
\end{equation}
Given point $A$ and slope $v_{\mathrm t}$, we have line $AD$ expressed as follows:
\begin{equation}
	AD: \  x = x_A + v_{\mathrm t} (t - t_A).
\end{equation}
From lines $DF$ and $AD$, we can obtain the intersection $D$ as follows:
\begin{equation}\label{equ:D}
D(A,F,v_{\mathrm t}, w):
\left\{
\begin{aligned}
t_D &= \dfrac{v_{\mathrm t} t_A - w t_F + (x_F - x_A)}{v_{\mathrm t} - w}, \\
x_D &= x_A + v_{\mathrm t} (t_D - t_A).
\end{aligned}
\right.
\end{equation}
Given point $D$ and slope $v_{\mathrm w}$, we have line $CD$ expressed as follows:
\begin{equation}
	CD: \  x = x_D + v_{\mathrm w} (t - t_D).
\end{equation}
From lines $CD$ and $CE$, we can obtain the intersection $C$ as follows:
\begin{equation}\label{equ:C}
C(D,E,v_{\mathrm w},w): 
\left\{
\begin{aligned}
t_C &= \dfrac{v_{\mathrm w} t_D - w t_E + (x_E - x_D)}{v_{\mathrm w} - w}, \\
x_C &= x_D + v_{\mathrm w} (t_C - t_D).
\end{aligned}
\right.
\end{equation}
Given point $C$ and slope $v^*$, we have line $BC$ expressed as follows:
\begin{equation}\label{equ:BC}
	BC: \  x = x_C + v^* (t - t_C).
\end{equation}
Given point $A$ and slope $v_{\mathrm t}$, we have line $AB$ expressed as follows:
\begin{equation}
	AB: \  x = x_A + v_{\mathrm t} (t - t_A).
\end{equation}
From lines $BC$ and $AB$, we can obtain the intersection $B$ as follows:
\begin{equation}\label{equ:B}
B(A,C,v_{\mathrm t},v^*): 
\left\{
\begin{aligned}
t_B &= \dfrac{v_{\mathrm t} t_A - v^{*} t_C + (x_C - x_A)}{v_{\mathrm t} - v^{*}}, \\
x_B &= x_A + v_{\mathrm t}(t_B - t_A).
\end{aligned}
\right.
\end{equation}
Finally, we have the JAD plan $A\xrightarrow{v_{\mathrm t}} B \xrightarrow{v^*} C$.

\subsection{Q2: What is the JAD plan? Part II: Point $A$}\label{sec:Q2}

In practice, the location and moment (i.e., point $A$) for dispatching a JAD vehicle may be determined by traffic managers. 
Specifically, a JAD vehicle initially stands by at a fixed roadside location or on-ramp, denoted by $x_A$. 
Once a downstream stop-and-go wave is confirmed to be developing, the JAD vehicle is dispatched at time $t_A$, subject to practical considerations such as the emergence of a sufficiently large traffic gap. 
However, this dispatch is also constrained by a feasible range of point $A$, and this subsection focuses on determining the feasible range of the starting point $A$.

\begin{figure}[htbp]
    \centering
    \includegraphics[width=\linewidth]{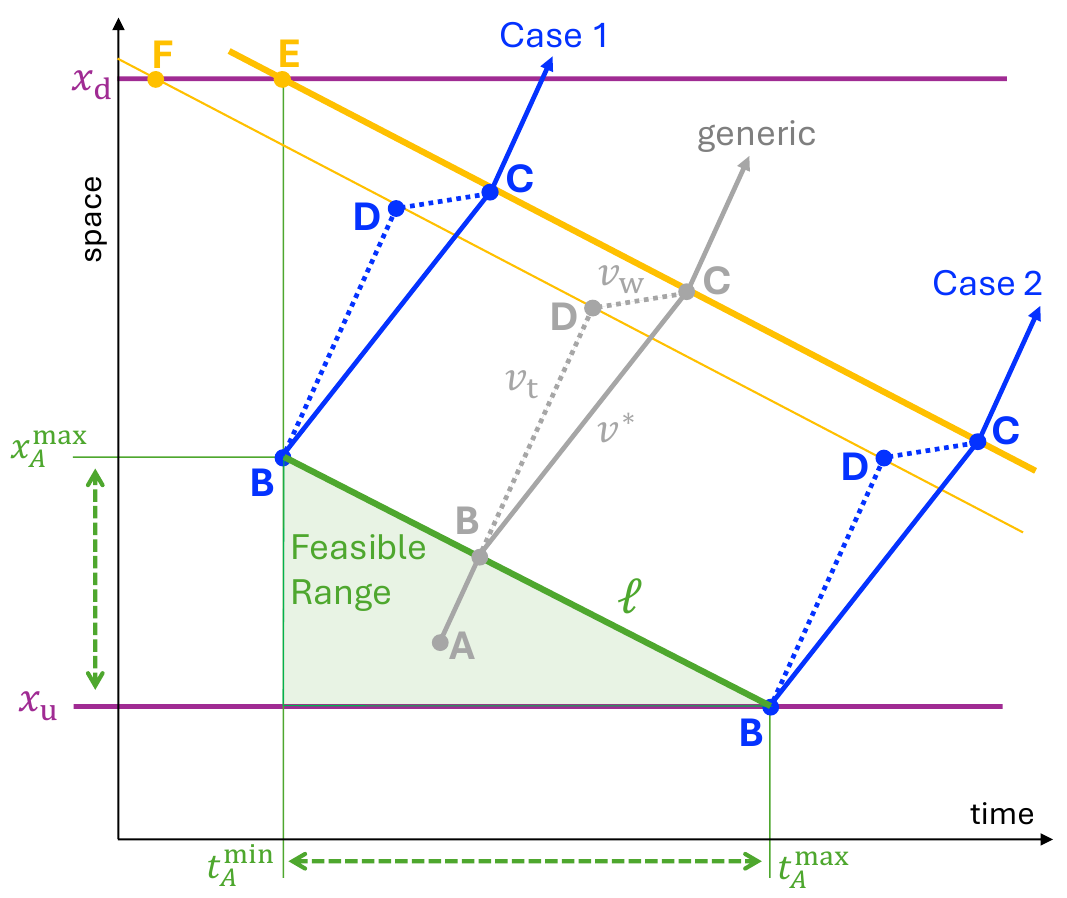}
    \caption{Feasible region of the dispatch time and location (point $A$) for a JAD vehicle.}
    \label{fig:jad_point_a}
\end{figure}

With the aid of Figure~\ref{fig:jad_point_a}, we analyze the feasible region of point $A$. 
We begin by temporarily ignoring point $A$\footnote{In Section~\ref{sec:Q2}, point $A$ is assumed to be known, and the corresponding JAD plan is then constructed. This sequence is consistent with the natural decision-making process in practice, where a JAD opportunity is first identified and subsequently planned. Here, however, to characterize the feasible region of point $A$, we do not assume that point $A$ is known a priori.} 
and consider the JAD Triangle $BCD$, whose sides $BC$, $BD$, and $CD$ have fixed slopes $v^{*}$, $v_{\mathrm t}$, and $v_{\mathrm w}$, respectively. 
Points $C$ and $D$ are constrained to lie on the parallel lines $CE$ and $DF$, both with slope $w$. 
Since the shape of the JAD Triangle $BCD$ is fixed and segment $CD$ is confined between these two parallel lines, the JAD Triangle $BCD$ can only undergo rigid translations along a direction with slope $w$, under which point $B$ traces a straight line, denoted by $\ell$, with slope $w$.

We have the following conditions.  
Point $C$ lies on line $CE$ in \eqref{equ:CE} and line $BC$ in \eqref{equ:BC}, giving
\begin{equation}
\label{equ:C_CE}
	x_C = x_E + w\,(t_C - t_E),
\end{equation}
\begin{equation}
\label{equ:C_BC}
	x_C = x_B + v^*\,(t_C - t_B).
\end{equation}
Point $D$ lies on line $DF$ in \eqref{equ:DF} and line $BD$, giving
\begin{equation}
\label{equ:D_DF}
	x_D = x_F + w\,(t_D - t_F),
\end{equation}
\begin{equation}
\label{equ:D_BD}
	x_D = x_B + v_{\mathrm t}\,(t_D - t_B).
\end{equation}
The slope of segment $CD$ is
\begin{equation}
\label{equ:CD_connected}
\frac{x_C - x_D}{t_C - t_D} = v_{\mathrm w},
\end{equation}
and, as shown in Figure~\ref{fig:jad_point_a}, the geometric relations among the points are as follows, 
\begin{equation}\label{equ:geometry}
\begin{split}
& t_F<t_E<t_D<t_C, \\
& x_D<x_C<x_E=x_F, \\
& 0\leq v_{\mathrm w} < v^* < v_{\mathrm t}, \\
& 0>w.\\
\end{split}
\end{equation}

Combining \eqref{equ:C_CE}--\eqref{equ:geometry}, we can derive the trajectory of point $B$ as
\begin{equation}\label{equ:ell}
x_B = x_E + w\, (t_B - t_E) +
\frac{w \Delta_{\mathrm w} (v_{\mathrm t}-v_{\mathrm w}) (w-v^*)}{(w-v_{\mathrm w})(v_{\mathrm t}-v^*)}.
\end{equation}

From a reactive perspective (i.e., without prediction), a JAD vehicle can be dispatched only after the formation of a stop-and-go wave is confirmed at time $t_E$ ($t_E > t_F$); see Case~1 in Figure~\ref{fig:jad_point_a}. 
The upstream detector located at $x_{\mathrm u}$ defines the farthest admissible standby location of the JAD vehicle; see Case~2 in Figure~\ref{fig:jad_point_a}. 
Therefore, the temporal and spatial lower bounds are given as follows:
\begin{equation}
\label{equ:B_bounds}
t_B \ge t_E, \qquad x_B \ge x_{\mathrm u}.
\end{equation}
The feasible region of point $B$ is then the portion of line $\ell$ that lies within these bounds.

Substituting the bounds in \eqref{equ:B_bounds} into line~$\ell$ in \eqref{equ:ell}, three intersection points are obtained, namely
\[
(t_{\min},\, x_{\max}), \qquad (t_{\min},\, x_{\min}), \qquad (t_{\max},\, x_{\min}).
\]
These three points form a triangle, which defines the feasible region of point~$A$, since point~$A$ is the backward extension of point~$B$ along the space-time direction.

\section{Theoretical analysis of the JAD}\label{sec:Analysis}

Segment $BC$ represents a JAD process with distance $J_{\mathrm{dis}}$ and duration $J_{\mathrm{dur}}$.
Given two parallel lines $CE$ in \eqref{equ:CE}, $\ell$ in \eqref{equ:ell}, and the geometric relations in \eqref{equ:geometry}, the spatial and temporal length of segment $BC$, which is an in-between segment with slope $v^*$, can be written as follows:
\begin{equation}\label{equ:jad_distance}
\begin{split}
	J_{\mathrm{dis}}(v^*, v_{\mathrm t}, \Delta_{\mathrm w}, w, v_{\mathrm w}) &=  x_C - x_B \\
		=&  - \frac{ v^* w \Delta_{\mathrm w} (v_{\mathrm t} - v_{\mathrm{w}}) (w - v^*) }{ (v^* - w)(w - v_{\mathrm{w}})(v_{\mathrm t} - v^*) },
\end{split}
\end{equation}
\begin{equation}\label{equ:jad_duration}
	J_{\mathrm{dur}}(\Delta_{\mathrm w}, w, v_{\mathrm t}, v_{\mathrm w}, v^*)
	= t_C - t_B
	= \frac{J_{\mathrm{dis}}}{v^*}.
\end{equation}
$J_{\mathrm{dis}}$ and $J_{\mathrm{dur}}$ are linearly related through the distance–time–speed relationship and are determined by five parameters, i.e., $v^*$, $v_{\mathrm t}$, $\Delta_{\mathrm w}$, $w$, and $v_{\mathrm w}$.
Therefore, analyzing the impact of these parameters on $J_{\mathrm{dis}}$ is sufficient to characterize the behavior of the JAD process, without loss of generality.

To facilitate the interpretation of the theoretical results, we summarize empirically plausible parameter ranges related to stop-and-go waves and JAD vehicles in Table~\ref{tab:parameter_ranges}. 
These values are consistent with real-world traffic observations and serve as practical references for the subsequent analysis of the impact of different factors on JAD implementation.

\begin{table*}[h]
\centering
\caption{Parameter ranges related to stop-and-go traffic waves and JAD strategies}
\label{tab:parameter_ranges}
\begin{tabular*}{\textwidth}{@{\extracolsep{\fill}} p{0.2\textwidth} p{0.04\textwidth} p{0.04\textwidth} p{0.04\textwidth} p{0.68\textwidth}}
\hline \noalign{\vskip 1mm}
\textbf{Parameter} & \textbf{Min} & \textbf{Max} & \textbf{Unit} & \textbf{Description} \vspace{0.6mm}\\
\hline \noalign{\vskip 1mm}
JAD speed ($v^*$) & 40 & 80 & km/h
& Desired JAD speed that may not trigger secondary waves, considering traffic flow stability \vspace{0.6mm}\\
Inflow traffic speed ($v_{\mathrm{t}}$) & 80 & 120 & km/h
& Speed of traffic approaching the stop-and-go traffic wave from upstream \vspace{0.5mm}\\
Wave temporal width ($\Delta_{\mathrm{w}}$) & 20 & 100 & s
& Temporal width of the stop-and-go traffic wave \vspace{0.6mm}\\
Wave speed ($w$) & $-20$ & $-10$ & km/h
& Wave speed of the stop-and-go traffic wave propagating upstream \vspace{0.6mm}\\
In-wave speed ($v_{\mathrm{w}}$) & 0 & 20 & km/h
& Traffic speed inside the moving queue associated with the stop-and-go traffic wave \\
\noalign{\vskip 1mm}
\hline
\end{tabular*}
\end{table*}

\subsection{Impact of JAD Speed}\label{sec:impact_jad_speed}

To analyze the impact of the JAD speed on the JAD process, we take the partial derivative of \eqref{equ:jad_distance} with respect to $v^*$:
\begin{equation}\label{equ:dJdis_dvstar}
	\frac{\partial J_{\mathrm{dis}}}{\partial v^*} 
	= \frac{w \, \Delta_{\mathrm w} \, (v_{\mathrm t} - v_{\mathrm w}) \, v_{\mathrm t}}{(w - v_{\mathrm w}) \, (v_{\mathrm t} - v^*)^2}.
\end{equation}
Considering the geometric relations in \eqref{equ:geometry}, we have $w-v_{\mathrm w}<0$, and it follows that
${\partial J_{\mathrm{dis}}}/{\partial v^*}>0$, which indicates that the JAD distance increases monotonically with the JAD speed.

\begin{figure}[htbp]
    \centering
    \includegraphics[width=\linewidth]{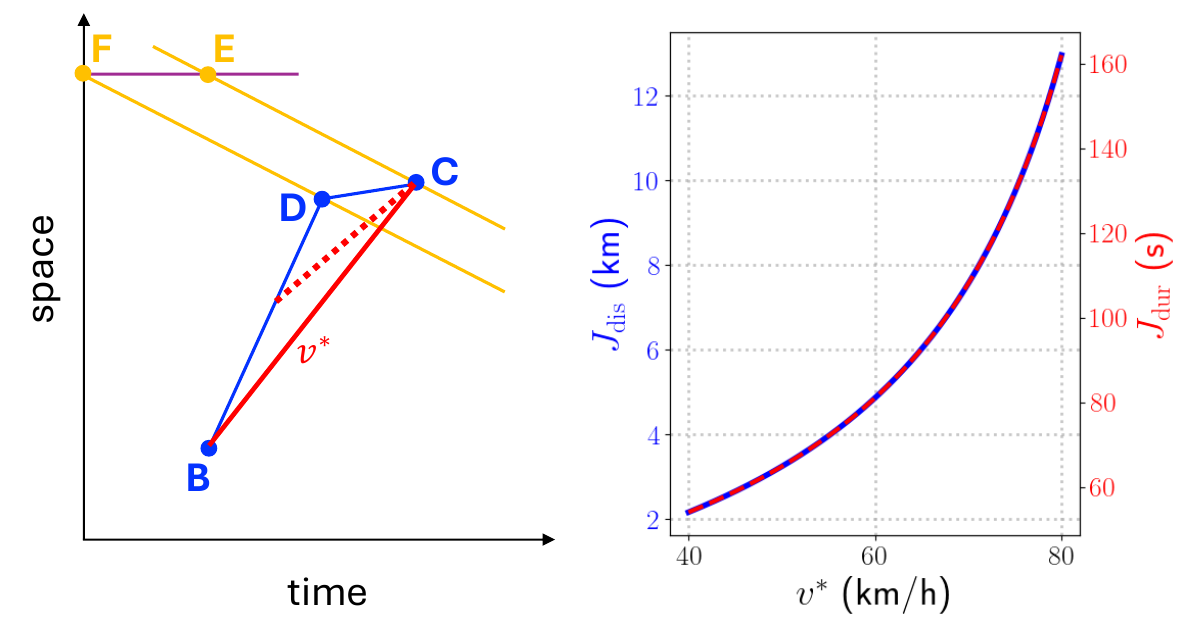}
    \caption{Impact of JAD speed $v^*$ on the JAD process ($J_{\mathrm{dis}}$ and $J_{\mathrm{dur}}$). Left: schematic diagram; right: numerical simulation results obtained by fixing all other parameters at their midpoint values in Table~\ref{tab:parameter_ranges}.}
    \label{fig:J_vs_v*}
\end{figure}

Physically, as illustrated in Figure~\ref{fig:J_vs_v*}, this implies that, with all other variables held constant, executing JAD at a higher target speed $v^*$ requires a longer distance to adequately interact with the stop-and-go traffic wave.

The nonlinear dependence of $J_{\mathrm{dis}}$ on $v^*$ is governed by the squared denominator $(v_{\mathrm t}-v^*)^2$. 
As $v^*$ approaches $v_{\mathrm t}$, the denominator tends to zero and the marginal increase $\partial J_{\mathrm{dis}}/\partial v^*$ grows rapidly. 
This indicates that the JAD distance becomes increasingly sensitive to the target speed at higher $v^*$, i.e., a convex dependence of the JAD distance on the target speed, also illustrated in Figure~\ref{fig:J_vs_v*}. 
Therefore, planning a high JAD speed requires a substantially longer execution distance.

A rough estimate using typical parameter ranges from Table~\ref{tab:parameter_ranges} shows that $\partial J_{\mathrm{dis}}/\partial v^*$ is on the order of $10^1\sim10^2$~m, which corresponds to a JAD duration on the order of $10^2\sim10^3~\mathrm{s}$. 
This implies that an increase of 1~km/h in the JAD speed may require an additional $10\sim100$~m of execution distance. 

The results of the numerical simulations shown in Figure~\ref{fig:J_vs_v*} indicate that $J_{\mathrm{dis}}$ ranges from approximately 2~km to more than 12~km, and $J_{\mathrm{dur}}$ ranges from approximately 50~s to more than 160~s, given moderate values of the other parameters.

\subsection{Impact of Inflow Traffic Speed}

Likewise, from \eqref{equ:jad_distance}, the partial derivative of the JAD distance with respect to $v_{\mathrm t}$ can be written as follows:
\begin{equation}\label{equ:dJdis_dvt}
\frac{\partial J_{\mathrm{dis}}}{\partial v_{\mathrm t}}
= \frac{v^* w \Delta_{\mathrm w} (w - v^*) (v^* - v_{\mathrm{w}})}{(v^* - w)(w - v_{\mathrm{w}})(v_{\mathrm t} - v^*)^2}.
\end{equation}
Considering the geometric relations in \eqref{equ:geometry}, we have $\frac{\partial J_{\mathrm{dis}}}{\partial v_{\mathrm t}} < 0$, indicating that a higher inflow traffic speed reduces the required JAD distance.

\begin{figure}[htbp]
    \centering
    \includegraphics[width=\linewidth]{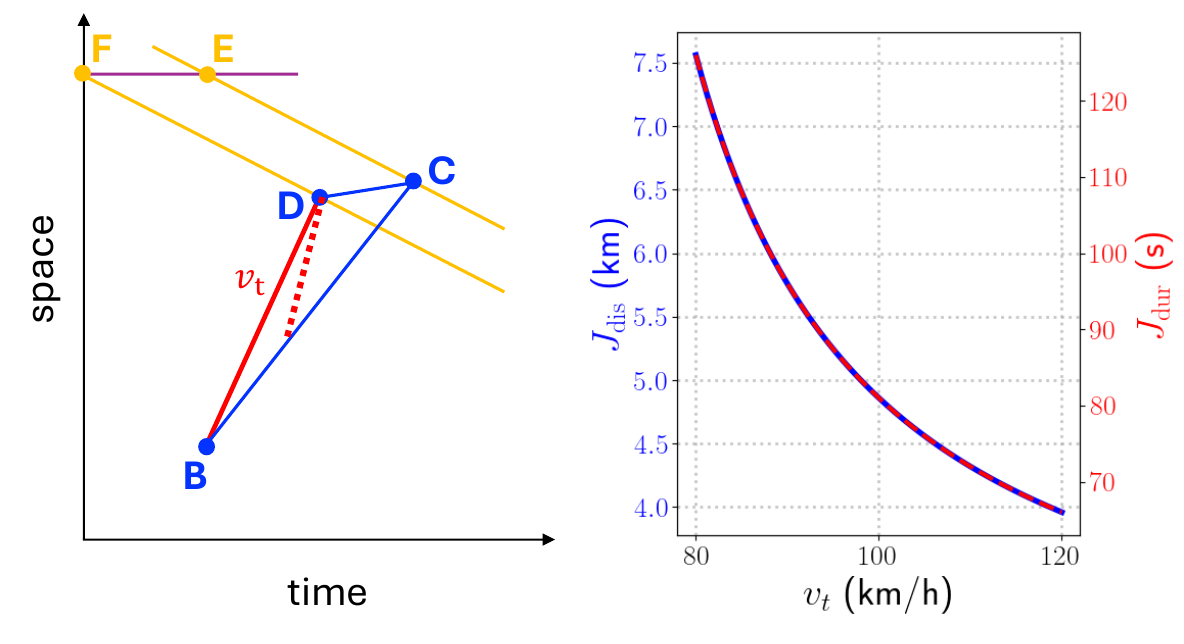}
    \caption{The impact of inflow traffic speed $v_t$ on the JAD process ($J_{\mathrm{dis}}$ and $J_{\mathrm{dur}}$).
    Left: schematic diagram; Right: results of numerical simulation, measured by keeping all other parameters at their midpoint values in Table~\ref{tab:parameter_ranges}.}
    \label{fig:J_vs_v_t}
\end{figure}

Physically, as illustrated in Figure~\ref{fig:J_vs_v_t}, a faster inflow (red dashed line) results in a shorter distance and a shorter duration of the JAD process, under the condition that all other variables are held constant. 
The nonlinear dependence of $J_{\mathrm{dis}}$ on $v_{\mathrm t}$ is dominated by the inverse-square term $(v_{\mathrm t}-v^*)^{-2}$ in \eqref{equ:dJdis_dvt}, implying that the marginal reduction diminishes as $v_{\mathrm t}$ increases.

A rough order-of-magnitude estimate based on typical parameter ranges in Table~\ref{tab:parameter_ranges} indicates that $\partial J_{\mathrm{dis}}/\partial v_{\mathrm t}$ is on the order of $10^1\sim10^2$~m, which corresponds to $J_{\mathrm{dur}}$ on the order of $10^1\sim10^2$~s. 
The results of the numerical simulations shown in Figure~\ref{fig:J_vs_v*} indicate that $J_{\mathrm{dis}}$ ranges from approximately 4 to 7.5~km and $J_{\mathrm{dur}}$ ranges from approximately 60 to 130~s, given moderate values of the other parameters.

\subsection{Impact of Wave Temporal Width}

From \eqref{equ:jad_distance}, the partial derivative of the JAD distance with respect to the wave width is
\begin{equation}\label{equ:dJdis_dEtFt}
\frac{\partial J_{\mathrm{dis}}}{\partial \Delta_{\mathrm w}} = \frac{v^* w (v_{\mathrm t} - v_{\mathrm w})}{(w - v_{\mathrm w})(v_{\mathrm t} - v^*)}.
\end{equation}
Considering the geometric relations in \eqref{equ:geometry}, we have $\frac{\partial J_{\mathrm{dis}}}{\partial \Delta_{\mathrm w}} > 0$, indicating that wider waves require longer JAD distances.

\begin{figure}[htbp]
    \centering
    \includegraphics[width=\linewidth]{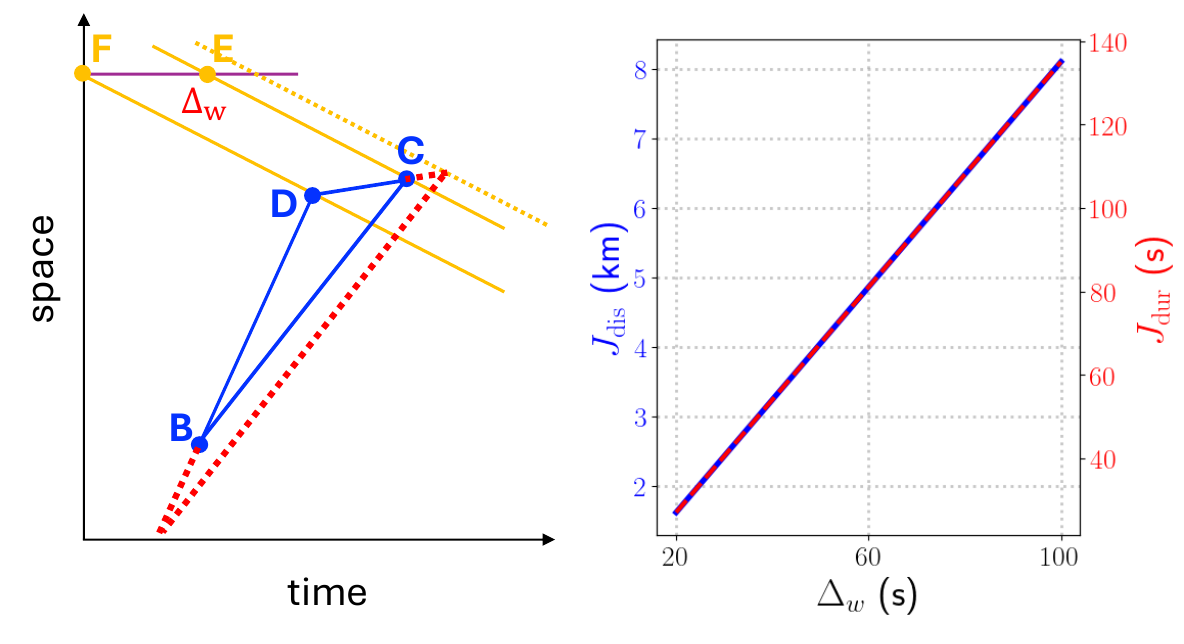}
    \caption{The impact of wave width $\Delta_{\mathrm w}$ on the JAD process ($J_{\mathrm{dis}}$ and $J_{\mathrm{dur}}$).
    Left: schematic diagram; Right: results of numerical simulation, measured by keeping all other parameters at their midpoint values in Table~\ref{tab:parameter_ranges}.}
    \label{fig:J_vs_Delta_w}
\end{figure}

Physically, a longer-lasting stop-and-go wave requires the JAD vehicle to travel a greater distance to fully interact with the moving queue (Figure~\ref{fig:J_vs_Delta_w}).

The dependence is linear in the wave width $\Delta_{\mathrm w}$, which is also confirmed by the numerical simulation shown in Figure~\ref{fig:J_vs_Delta_w}. 
As can be seen from the theoretical analysis in the left subfigure of Figure~\ref{fig:J_vs_Delta_w}, an increase in $\Delta_{\mathrm w}$ leads to a proportional expansion of the JAD Triangle $BCD$.

Increasing the wave duration from its lower to upper bound in Table~\ref{tab:parameter_ranges} typically increases $J_{\mathrm{dis}}$ by one to two orders of magnitude ($10^1 \sim 10^2$~m), corresponding to $J_{\mathrm{dur}}$ on the order of $10^1 \sim 10^2$~s. 
The simulation results in Figure~\ref{fig:J_vs_Delta_w} further indicate that, for moderate values of the other parameters, $J_{\mathrm{dis}}$ ranges from approximately 2 to 8~km, and $J_{\mathrm{dur}}$ ranges from approximately 30 to 140~s.

\subsection{Impact of Wave Speed}

\begin{figure}[htbp]
    \centering
    \includegraphics[width=\linewidth]{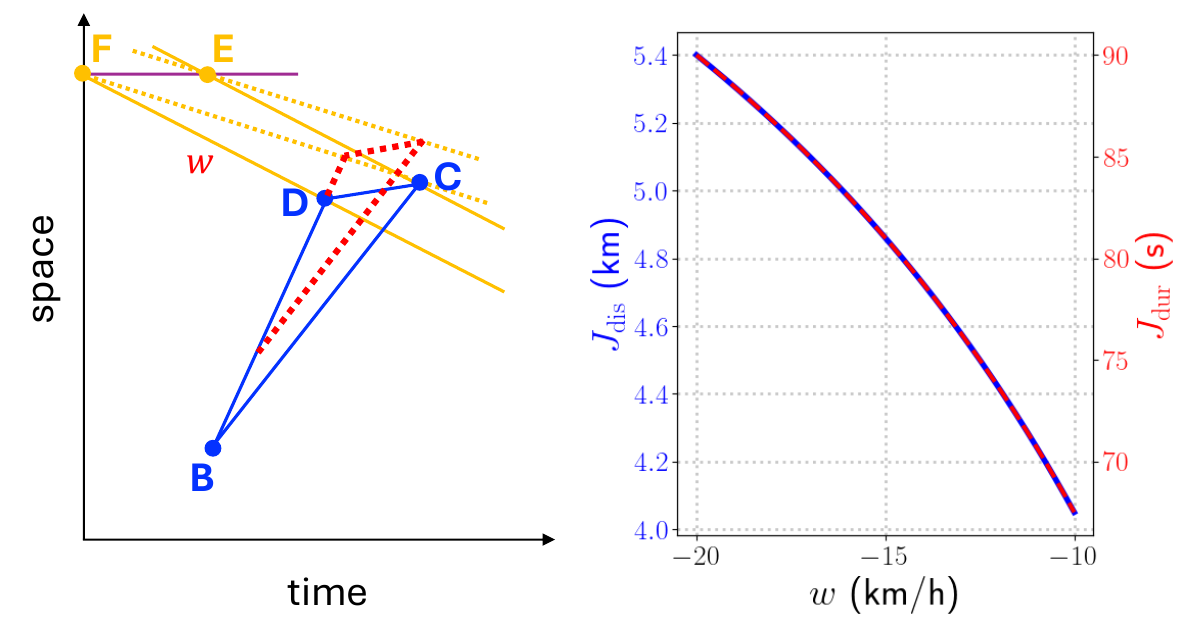}
    \caption{The impact of wave speed $w$ on the JAD process ($J_{\mathrm{dis}}$ and $J_{\mathrm{dur}}$).
    Left: schematic diagram; Right: results of numerical simulation, measured by keeping all other parameters at their midpoint values in Table~\ref{tab:parameter_ranges}.}
    \label{fig:J_vs_w}
\end{figure}

From \eqref{equ:jad_distance}, the partial derivative of the JAD distance with respect to the wave speed is given by
\begin{equation}\label{equ:dJdis_dw}
\frac{\partial J_{\mathrm{dis}}}{\partial w}
= - \frac{\Delta_{\mathrm w} \, v^* \, v_{\mathrm w} (v_{\mathrm t} - v_{\mathrm w})}
{(v_{\mathrm t} - v^*)(w - v_{\mathrm w})^2}.
\end{equation}
Considering the geometric relations in \eqref{equ:geometry}, all terms in the denominator are positive, and the numerator is also positive except for the leading minus sign. Hence, $\partial J_{\mathrm{dis}}/\partial w < 0$. Since $w < 0$, an increase in $w$ (i.e., a reduction in its magnitude) corresponds to a slower propagation of the stop-and-go wave. Therefore, a slower wave speed leads to a shorter required JAD distance and duration (Figure~\ref{fig:J_vs_w}).

As shown in \eqref{equ:dJdis_dw}, the sensitivity of $J_{\mathrm{dis}}$ to $w$ scales with $(w - v_{\mathrm w})^{-2}$, implying a rapid change in the required JAD distance as the assumed wave speed approaches the actual propagation speed of the stop-and-go wave. This nonlinear amplification effect is clearly illustrated by the numerical results in Figure~\ref{fig:J_vs_w}.

Varying the wave speed within the range reported in Table~\ref{tab:parameter_ranges} typically leads to changes in $J_{\mathrm{dis}}$ on the order of $10^2 \sim 10^3$~m, and in $J_{\mathrm{dur}}$ on the order of $10^1 \sim 10^2$~s. 
The simulation results in Figure~\ref{fig:J_vs_w} further show that, for moderate values of the other parameters, $J_{\mathrm{dis}}$ ranges from approximately 4 to 5.4~km, and $J_{\mathrm{dur}}$ ranges from approximately 65 to 90~s. 
Compared with the ranges resulting from other parameters, the impact of wave speed on the JAD process is relatively small.

\subsection{Impact of In-Wave Traffic Speed}

From \eqref{equ:jad_distance}, the partial derivative of the JAD distance with respect to $v_{\mathrm w}$ is
\begin{equation}\label{equ:dJdis_dvw}
\frac{\partial J_{\mathrm{dis}}}{\partial v_{\mathrm w}} = \frac{v^* w \Delta_{\mathrm w} (v_{\mathrm t} - w)}{(v_{\mathrm t} - v^*) (w - v_{\mathrm w})^2}.
\end{equation}
Considering the geometric relations in \eqref{equ:geometry}, we have that $\partial J_{\mathrm{dis}}/\partial v_{\mathrm w} < 0$, indicating that a higher in-wave traffic speed reduces the required JAD distance.

Physically, when the vehicles inside the stop-and-go wave move faster, the JAD Triangle becomes smaller, and the JAD process is shortened accordingly. 
This means that, compared with a stop-and-go wave, a slower moving-and-go wave is easier to mitigate, which is consistent with common sense.

\begin{figure}[htbp]
    \centering
    \includegraphics[width=\linewidth]{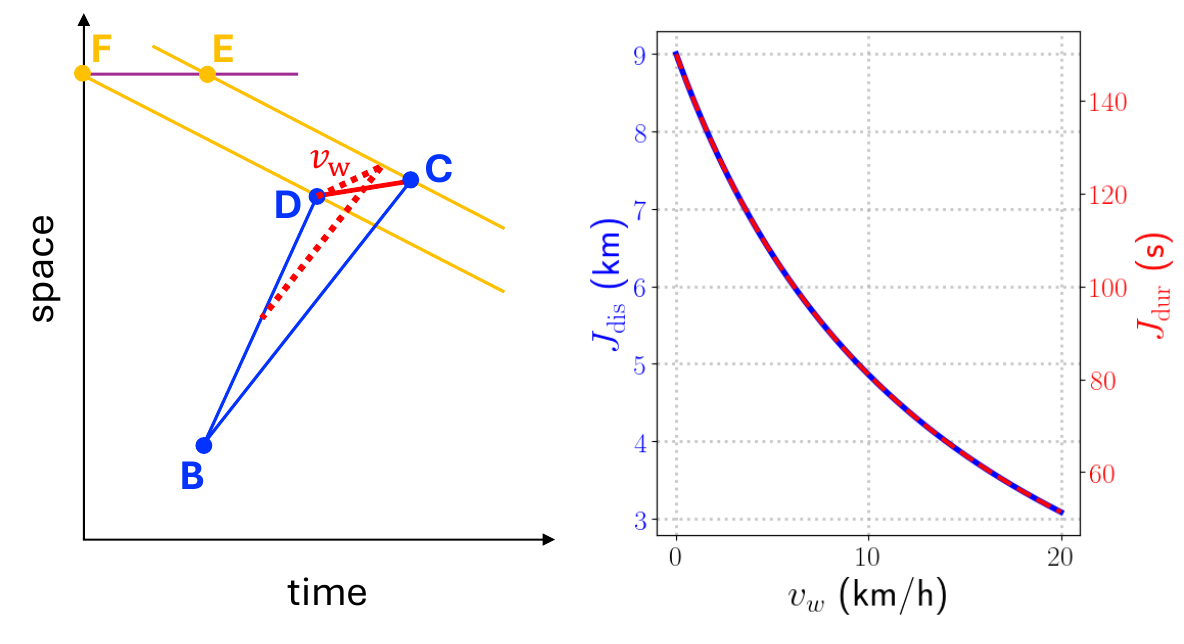}
    \caption{The impact of in-wave traffic speed $v_\text{w}$ on the JAD process ($J_{\mathrm{dis}}$ and $J_{\mathrm{dur}}$).
    Left: schematic diagram; Right: results of numerical simulation, measured by keeping all other parameters at their midpoint values in Table~\ref{tab:parameter_ranges}.}
    \label{fig:J_vs_v_w}
\end{figure}

The nonlinear dependence on $v_{\mathrm w}$ is reflected in the squared term $(w - v_{\mathrm w})^2$ in the denominator. 
For typical values of $v_{\mathrm w}$ in the range 0$\sim$20~km/h, variations in $J_{\mathrm{dis}}$ are moderate on the order of $10^1 \sim 10^2$~m, corresponding to $J_{\mathrm{dur}}$ on the order of $10^2 \sim 10^3$~s. 
The simulation results in Figure~\ref{fig:J_vs_v_w} further show that, for moderate values of the other parameters, the range of $J_{\mathrm{dis}}$ is between 3~km and 9~km, and the range of $J_{\mathrm{dur}}$ is between 60~s and 150~s.

\subsection{Summary of Parameter Sensitivity}

Based on the preceding theoretical analysis, the sensitivity of the JAD implementation to key parameters can be summarized as follows, as shown in Table~\ref{tab:sensitivity_summary}.
\begin{itemize}
    \item The impacts of JAD speed $v^*$ and wave width $\Delta_{\mathrm{w}}$ on $J_{\mathrm{dis}}$ and $J_{\mathrm{dur}}$ are positive, i.e., higher $v^*$ and larger $\Delta_{\mathrm{w}}$ lead to a longer JAD process, while keeping all other parameters fixed.
    
    \item The impacts of inflow traffic speed $v_{\mathrm t}$, wave speed $w$, and in-wave traffic speed $v_{\mathrm{w}}$ on $J_{\mathrm{dis}}$ and $J_{\mathrm{dur}}$ are negative, indicating that a lower $v_{\mathrm t}$, a faster propagating wave (smaller $|w|$), and lower $v_{\mathrm{w}}$ require a shorter JAD process.
    
    \item All the impacts are nonlinear, except for that of the wave width $\Delta_{\mathrm{w}}$, which is linear.
    
    \item Considering moderate values of the other parameters, the relative influence of each parameter on $J_{\mathrm{dis}}$ and $J_{\mathrm{dur}}$, from strongest to weakest, is:
    $$
    v^* > \Delta_{\mathrm{w}} > v_{\mathrm{w}} > v_t > w,
    $$
    and the corresponding ranges are listed in Table~\ref{tab:sensitivity_summary}.
\end{itemize}

\begin{table}[t]
\centering
\caption{Summary of parameter influence on JAD implementation}
\label{tab:sensitivity_summary}
{%
\begin{tabular}{p{1cm} p{2.4cm} p{0.9cm} p{1.2cm} p{1.2cm}}
\hline \noalign{\vskip 1mm}
\makecell[l]{\textbf{Impact} \\ \textbf{Ranking}}  & \textbf{Parameter} & \textbf{Direction} & $\boldsymbol{J_{\mathrm{dis}}}$ (km) & $\boldsymbol{J_{\mathrm{dur}}}$ (s)  \vspace{0.6mm}\\
\hline \noalign{\vskip 1mm}
\bf{1} & JAD speed ($v^*$) & $+$  & [2, 13]  & [50, 160]  \vspace{0.6mm}\\
\bf{2} & Wave width ($\Delta_{\mathrm w}$) & $+$ & [1, 8] & [30, 140]  \vspace{0.6mm}\\
\bf{3} & Wave speed ($|w|$) & $+$ & [4, 5.4] & [65, 90]   \vspace{0.6mm}\\
\bf{4} & Inflow speed ($v_{\mathrm t}$) & $-$ & [4, 7.5]  & [60, 130]   \vspace{0.6mm}\\
\bf{5} & In-wave speed ($v_{\mathrm w}$) & $-$ & [3, 9] & [60, 150]\\
\noalign{\vskip 1mm}
\hline 
\end{tabular}%
}
\parbox{\columnwidth}{\scriptsize
\vspace{2mm}
Note: ``$+$'' (``$-$'') indicates that the parameter increases (decreases) the corresponding JAD quantity.}
\end{table}

Referring to real-world scenarios, such temporal and spatial requirements are within the operational capabilities of JAD vehicles, including police cars exhibiting the observed swerving behavior, implying that the proposed JAD strategy is feasible and implementable under typical traffic conditions.

\section{Simulation-based Analysis and Practical Setting Demonstration}\label{sec:Simulation}

\subsection{Simulation Scenario}

It is known that speed variance between lanes is usually small, as lane-changing behavior tends to smooth out inter-lane differences \cite{PhysRevE,Treiber2011a,VRANKEN}. 
Therefore, single-lane traffic or the average across all lanes is representative of multi-lane traffic. 
Without loss of generality, we simulate traffic on a single-lane straight freeway. 
SUMO is employed to ensure reproducibility. 
The road length is 8 km, with a speed limit of 90 km/h. 
The simulation runs for 1600~s with a time step of 1~s. 
Vehicles enter the network at the freeway entrance every 1.5~s\footnote{Such persistently high demand is effective in ensuring the continuous growth of a stop-and-go wave. We also experimented with more stochastic demand patterns but failed to produce a self-sustaining wave.} with an initial speed of 54 km/h, and they quickly accelerate to their maximum speed. 
The Krauss car-following model embedded in SUMO is employed, and the main parameters are summarized in Table~\ref{tab:Krauss}.

\begin{table}[htbp]
\caption{Main Parameters of the Krauss Model}
\begin{tabular}{p{0.4\columnwidth} p{0.21\columnwidth} p{0.21\columnwidth}}
\hline \noalign{\vskip 1mm}
\bf Parameter & \bf Value & \bf Unit \\
\hline \noalign{\vskip 1mm}
Max Acceleration        & 2.0  & m/s² \vspace{0.5mm}\\
Max Deceleration        & 4.5  & m/s² \vspace{0.5mm}\\
Desired Reaction Time   & 1.2  & s    \vspace{0.5mm}\\
Reaction Time Variation & 1.2  & s    \vspace{0.5mm}\\
Acceleration Variation  & 0.95 & –    \vspace{0.5mm}\\
Deceleration Variation  & 0.95 & –    \vspace{0.5mm}\\
Driver Randomness       & 0.95 & –    \vspace{0.5mm}\\
Vehicle Length          & 5    & m    \vspace{0.5mm}\\
Minimum Gap             & 1.5  & m    \vspace{0.5mm}\\
Max Speed               & 25   & m/s  \vspace{0.5mm}\\
Speed Factor            & 1.0  & –    \vspace{0.5mm}\\
Speed Deviation         & 0.0  & –    \vspace{0.5mm}\\
\hline 
\end{tabular}\centering\label{tab:Krauss}
\end{table}

\begin{figure}[htbp]
    \centering
    \includegraphics[width=0.9\linewidth]{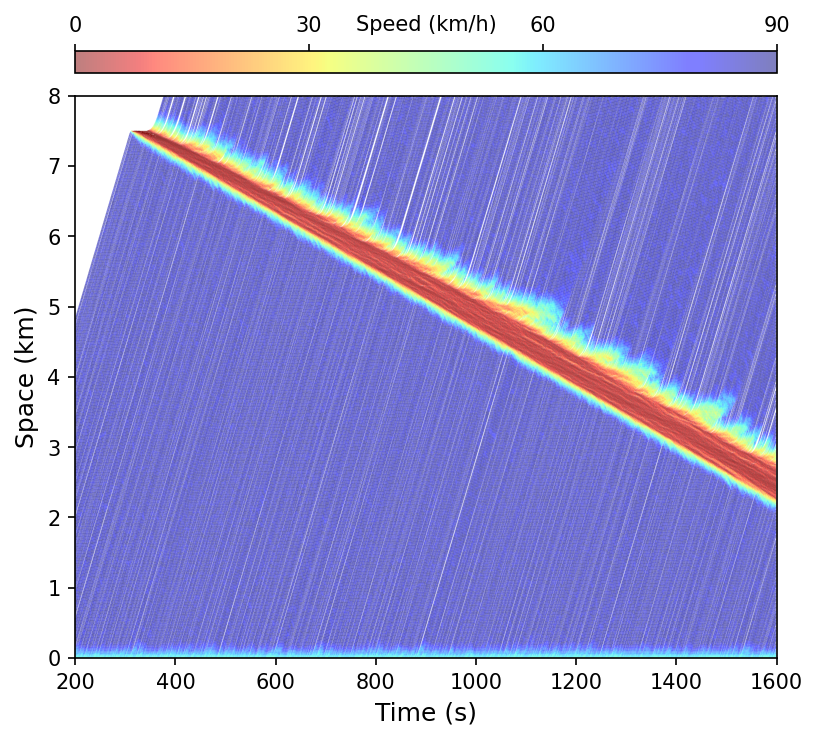}
    \caption{SUMO simulated self-sustaining stop-and-go wave.}
    \label{fig:base_trajectory}
\end{figure}

To test the JAD strategy, a disturbance is artificially introduced by stopping the first vehicle in the simulation 500~m before the end of the road for 30~s. 
This generates a clear and self-sustaining stop-and-go wave that propagates upstream and does not dissipate on its own (Figure~\ref{fig:base_trajectory}).

\subsection{Estimation of JAD Parameters}

Through the simulation scenario, we discuss and demonstrate the practical estimation of several key parameters.

\subsubsection{JAD Speed ($v^*$)}

As discussed in Section~\ref{sec:jad_speed}, traffic stability is the key to estimating a reliable JAD speed that will not trigger secondary waves. 
For the purpose of demonstrating the JAD strategy, we determine here, through simulation, a JAD speed that avoids triggering secondary waves. 
As noted previously, more empirical studies and results are still needed to advance the understanding of traffic stability in practice.

\begin{figure}[htbp]
    \centering
    \includegraphics[width=\linewidth]{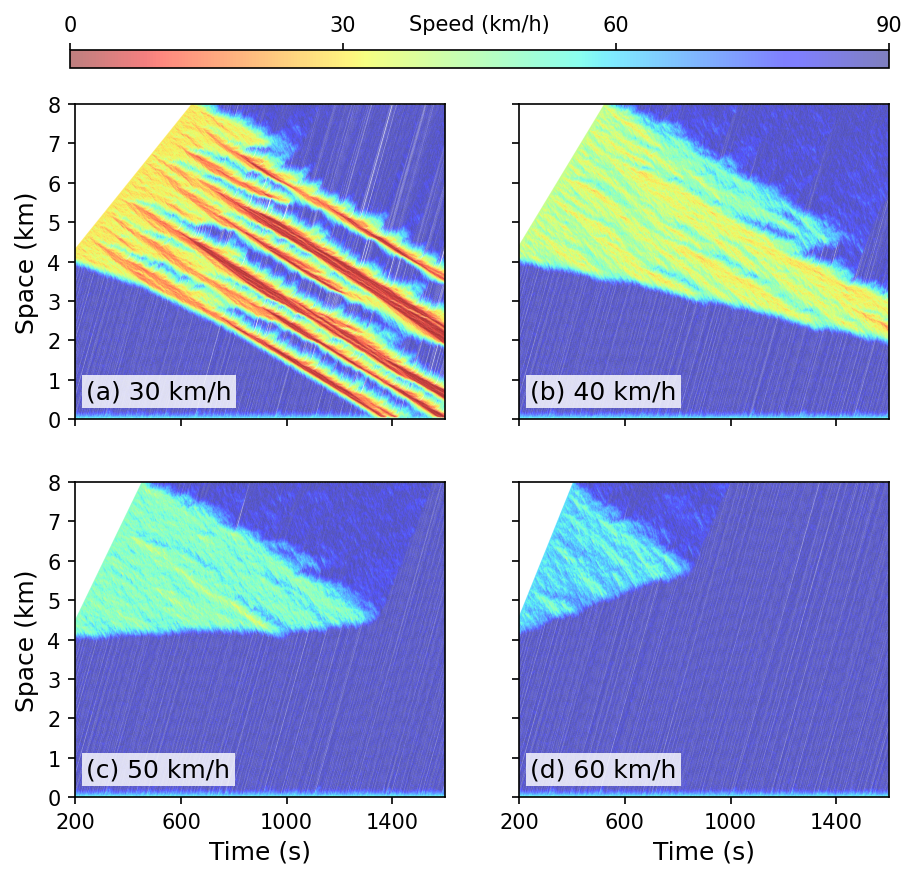}
    \caption{Comparison of traffic breakdown under moving bottlenecks with different speeds. 
    (a)(b) traffic breakdown;
    (c)(d) no breakdown.}
    \label{fig:stability_trajectory}
\end{figure}

We reduce the speed of the first vehicle to 30, 40, 50, and 60 km/h, respectively, and simulate each scenario to observe which speeds do not trigger secondary waves. 
As shown in Figure~\ref{fig:stability_trajectory}, speeds of 30 and 40 km/h result in traffic breakdown and are therefore unsuitable as JAD speeds. 
In contrast, speeds of 50 km/h or higher can serve as reliable JAD speeds. 
However, as discussed in Section~\ref{sec:impact_jad_speed}, increasing the JAD speed also increases the time and space required to execute the JAD task, implying a longer operational duration.

Therefore, in this simulation scenario, we set the JAD speed to $v^* = 55$ km/h.

\subsubsection{Wave temporal width ($\Delta_{\mathrm{w}}$)}

To mimic real-world conditions, we estimate the growth of a stop-and-go wave from the perspective of a stationary roadside detector.

Typically, a stationary detector monitors traffic and measures the speeds of passing vehicles in real time. 
Therefore, we assume that a stop-and-go wave has formed when the observed speed falls below a given threshold and remains low for a certain period of time. 
This approach is feasible because the focus of this study is an isolated stop-and-go wave. 
In contrast, for periodic stop-and-go waves generated by fixed bottlenecks, this method may not be suitable, and the proposed JAD strategy is not designed for such traffic conditions.

\begin{figure}[htbp]
    \centering
    \includegraphics[width=0.9\linewidth]{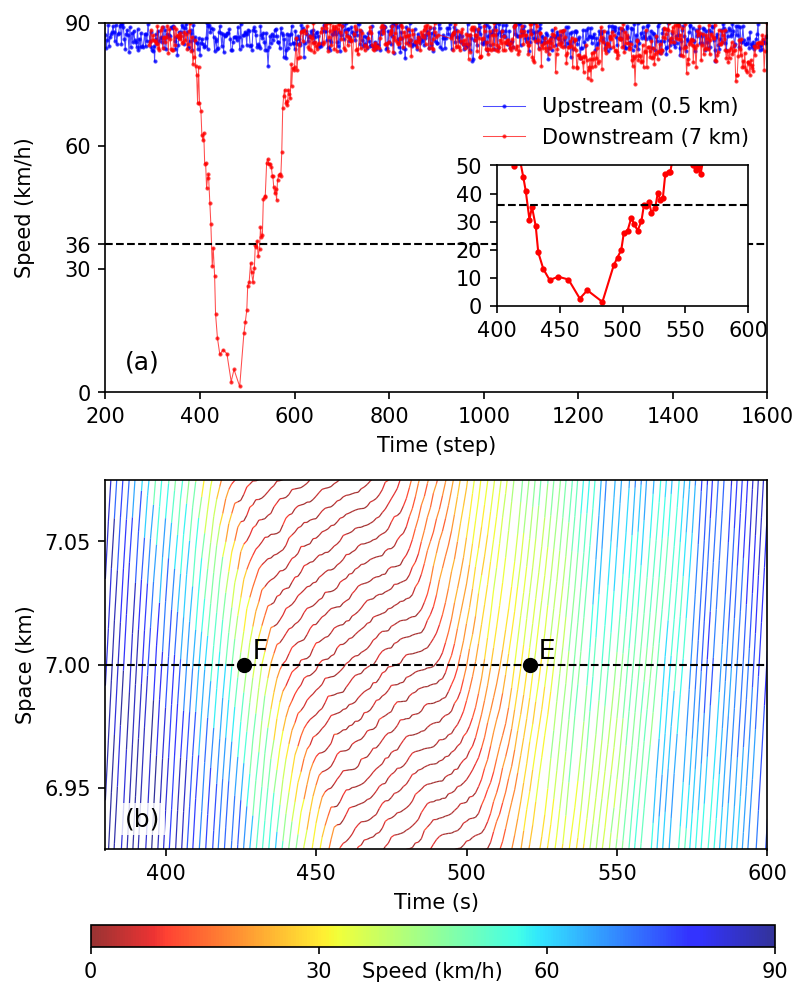}
    \caption{Estimation of wave temporal width using a stationary roadside detector.
    (a) Time series of detected vehicle speed;
    (b) Time-space trajectories around the wave.}
    \label{fig:jad_detector}
\end{figure}

In this scenario, we set the threshold to 36~km/h and the holding period to 30~s. 
As shown in Figure~\ref{fig:jad_detector}, the resulting time-space points $F$ and $E$ effectively capture the stop-and-go wave. 
It is also noted that, due to the asymmetric vehicle dynamics when traversing a stop-and-go wave \cite{Chen2012i,Li2013c,Park2019}, the recovery process from a slow-moving queue is longer than the deceleration process. 
As will be demonstrated later in Section~\ref{sec:simulation_result} through simulation results, a slight overestimation of the wave speed, while potentially sacrificing some efficiency, not only does not compromise the objective of suppressing stop-and-go wave propagation but may also improve the robustness of the JAD strategy.
If one seeks to avoid generating such an artificial gap in traffic flow, a more advanced strategy, such as the two-step strategy proposed in \cite{JAD2017}, can be adopted. However, it should be noted that such methods typically require more real-time traffic information to support fine-grained control.

\subsubsection{In-wave Speed ($v_{\mathrm{w}}$)}

From the practical perspective of a stationary roadside detector, the speed variation through a stop-and-go wave exhibits a gradual decrease followed by a gradual increase (also illustrated in Figure~\ref{fig:jad_detector}). 
As a result, a stationary roadside detector may be unable to accurately capture the in-wave speed.

Here, we define the in-wave speed $v_{\mathrm{w}}$ as the average speed between $t_F$ and $t_E$. 
If the minimum value were used instead, $v_{\mathrm{w}}$ could be underestimated.

\subsubsection{Wave Speed ($w$)}

Although it is difficult to determine the specific speed of an upcoming wave using a single stationary roadside detector, it can be reasonably treated as prior knowledge in practice for the following two reasons. 
First, global empirical studies have shown that wave speeds typically range from -20 to -10 km/h, indicating a relatively narrow range. 
Second, based on over one year of time-space trajectory data from the I-24 MOTION dataset \cite{JI2025105313}, it can be observed that waves occurring at the same locations tend to propagate at similar speeds. 
Therefore, the wave speed can be treated as prior knowledge.

In the simulation, we set the wave speed to $w = -15$ km/h.

\subsubsection{Inflow Speed $v_{\mathrm{t}}$}
The determination of the inflow speed in practice is straightforward. 
It can be obtained either from an upstream stationary roadside detector or from observations made by the driver of the JAD vehicle.

Here, we set the inflow speed $v_{\mathrm{t}}$ equal to that of the potential leading vehicle, i.e., we assume that the driver of the JAD vehicle can observe this speed.

\subsubsection{The Role of the Upstream Detector}

The upstream detector provides two key pieces of information. 
First, it measures the inflow rate, which is critical for determining whether the current stop-and-go wave is self-sustaining. 
If the inflow is observed to decrease, dispatching a JAD vehicle may not be necessary, as the stop-and-go wave may dissipate naturally. 
Second, it provides the inflow speed, which is used to determine $v_{\mathrm{t}}$ in the JAD plan, as described above.

\subsubsection{Opportunity of Dispatching a JAD Vehicle}

Inserting a new vehicle into traffic requires sufficient space. 
Here, we propose monitoring the time headway at the standby location of the JAD vehicle. 
When a specified condition is satisfied, we assume that it is feasible for the JAD vehicle to merge into traffic.

In the police-car swerving scenario, the police officer is expected to identify an appropriate gap for merging. 
If merging from an on-ramp, the overall process is similar to typical freeway merging behavior in daily traffic. 
With the siren activated, the required gap may be even smaller than under normal conditions.

In this study, we assume that a JAD vehicle can merge into traffic when the time headway exceeds 3~s. 
In simulations, we also tested smaller thresholds, such as 2~s, and found that they do not significantly disturb traffic. 
One reason is that, at point $A$, the JAD vehicle can merge at a relatively high speed (e.g., when merging from an on-ramp), close to the prevailing traffic speed or the speed of the potential leading vehicle.

\begin{figure}[htbp]
    \centering
    \includegraphics[width=0.9\linewidth]{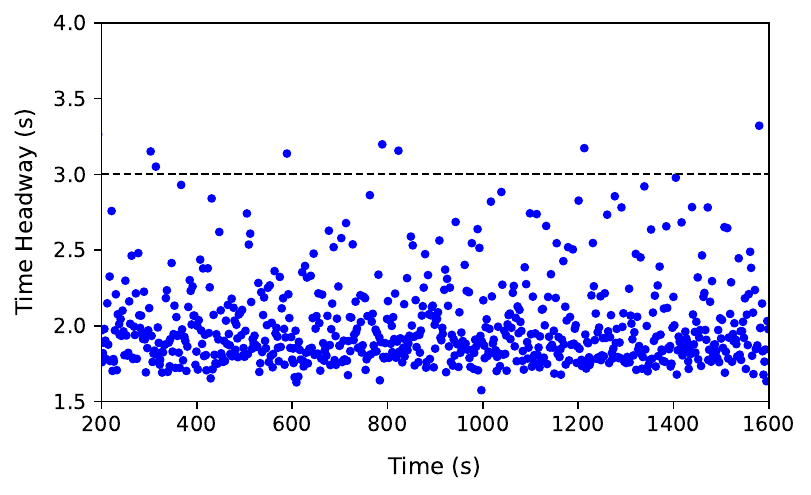}
    \caption{Time headways at the JAD vehicle's standby location.}
    \label{fig:jad_headway}
\end{figure}

\subsection{Simulation Results}\label{sec:simulation_result}

To moderate multi-lane traffic, swerving behavior is required. 
Since such behavior has already been observed in the real world (Figure~\ref{fig:swerving}), it is unnecessary to simulate or verify its feasibility. 
Therefore, we focus on evaluating the stop-and-go wave mitigation performance of the JAD strategy rather than the specific swerving maneuvers of the JAD vehicle.

\begin{figure}[htbp]
    \centering
    \includegraphics[width=0.9\linewidth]{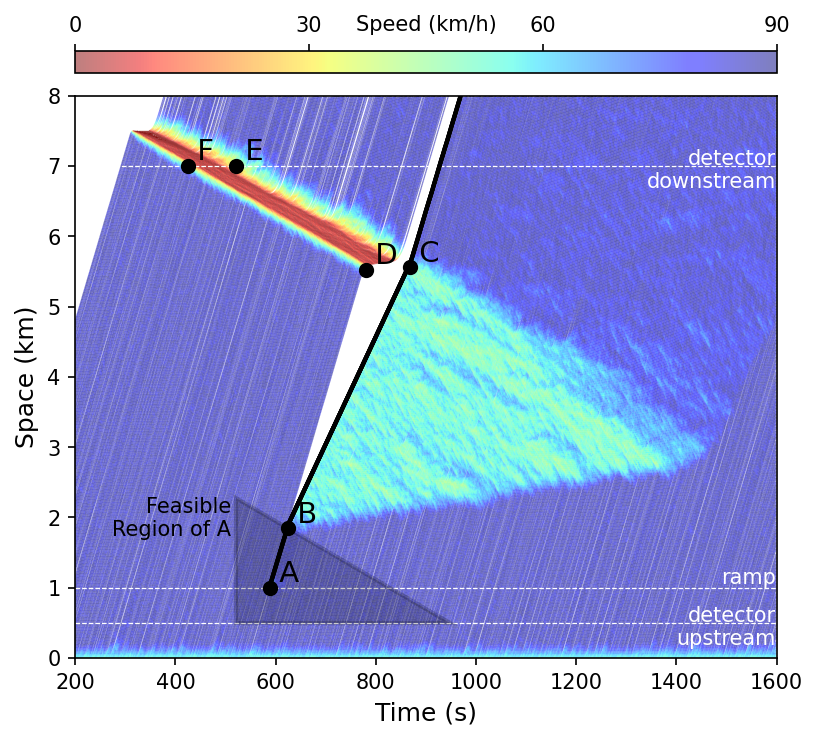}
    \caption{A stop-and-go wave is successfully suppressed using a JAD strategy with a JAD speed of 55 km/h.}
    \label{fig:jad_trajectory}
\end{figure}

Figure~\ref{fig:jad_trajectory} presents the simulation results. 
It illustrates the JAD plan (points $A$, $B$, $C$, $D$, $E$, and $F$) as well as the feasible region of point $A$. 
The results clearly show that the proposed JAD strategy successfully suppresses the propagation of the stop-and-go wave, with no secondary waves being triggered.

Meanwhile, we observe that a gap appears after the JAD vehicle completes its operation at point $C$. 
From the perspective of suppressing stop-and-go wave propagation, having point $C$ located slightly farther from point $D$ does not pose a significant issue. 
This is a positive indication, as a conservative (slightly overestimated) choice of point $C$ does not degrade performance, implying good robustness of the JAD strategy.

The results could be further improved through parameter fine-tuning. 
For example, a more accurate estimation of the wave width could enhance efficiency by enabling a better prediction of the ending point $C$ of the JAD strategy. 
However, this refinement may not be easy to implement in practice, because traffic conditions cannot be perfectly replicated and obtaining exact values is challenging.


Note that it is {\it not necessary} to compare travel time or emissions with the baseline in Figure~\ref{fig:base_trajectory}, since once the stop-and-go wave is suppressed, the following vehicles naturally experience shorter travel times and lower emissions under smoother traffic conditions. 
This effect becomes more pronounced when a large spatiotemporal scale is considered. 
Therefore, such comparisons are redundant, as improvements in travel efficiency and emissions are inherent outcomes of successful wave suppression rather than independent performance measures.

\begin{figure}[htbp]
    \centering
    \includegraphics[width=0.9\linewidth]{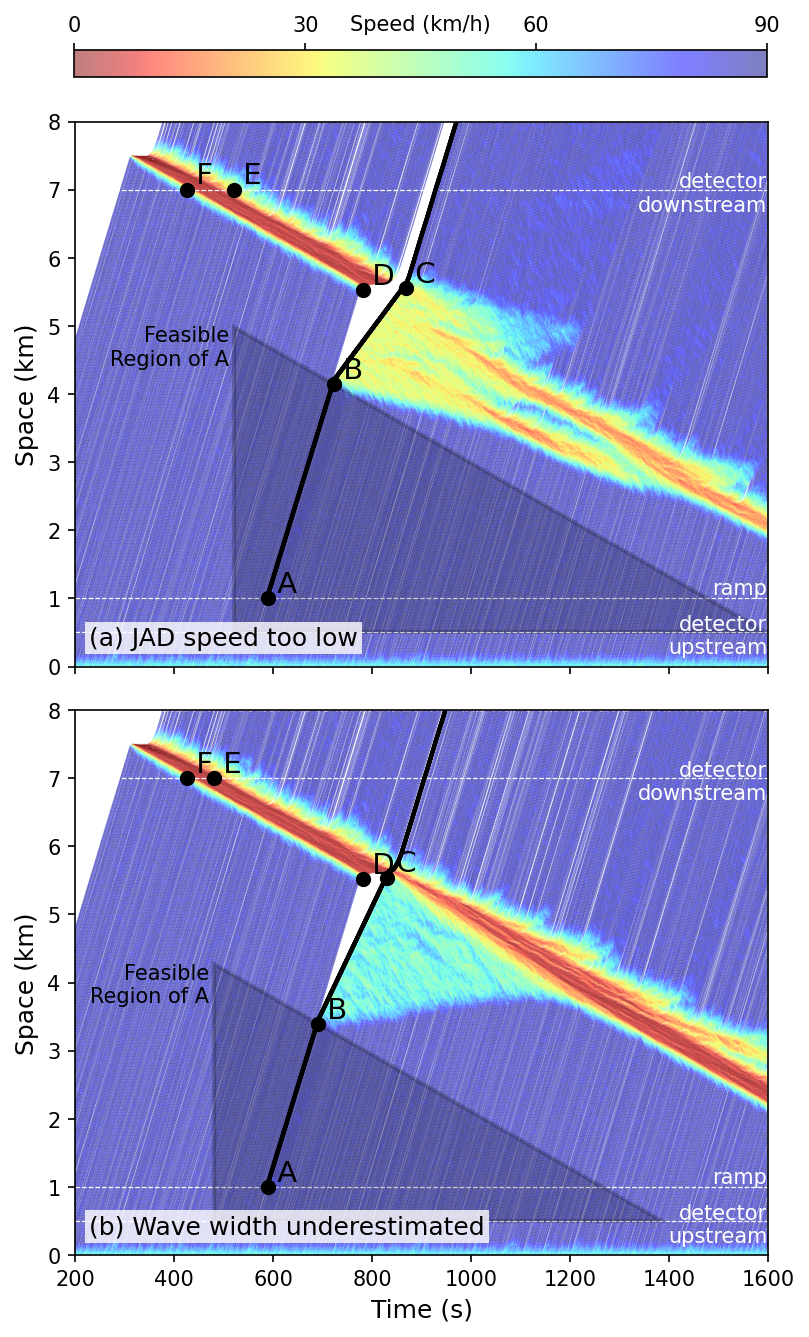}
    \caption{A JAD strategy that fails to suppress a stop-and-go wave. 
    (a) The JAD speed $v^* = 35$ km/h is too low to prevent traffic breakdown; 
    (b) point $C$ is located too close to point $D$, leading to insufficient suppression of the stop-and-go wave.}
    \label{fig:jad_trajectory_failed}
\end{figure}

Figure~\ref{fig:jad_trajectory_failed} presents two typical cases in which the JAD vehicle fails to suppress the stop-and-go wave, as secondary waves are triggered. 
In Figure~\ref{fig:jad_trajectory_failed}(a), the JAD speed is set to $v^* = 35$ km/h. As indicated by the stability test in Fig.~\ref{fig:stability_trajectory}, such a low speed is insufficient to prevent traffic breakdown. 
In Figure~\ref{fig:jad_trajectory_failed}(b), the wave width is (intentionally) underestimated, and the JAD vehicle fails to avoid being captured by the stop-and-go wave. As a result, the stop-and-go wave penetrates the JAD trajectory and continues to propagate upstream.

\section{Conclusion}\label{sec:Conclusion}

Inspired by real-world observations of police-car swerving behavior, this paper proposes a practical JAD strategy that transforms such real-world observed behavior into a JAD maneuver capable of suppressing the propagation of an isolated stop-and-go wave. 
Five key parameters that significantly affect the JAD strategy, namely, {\it JAD speed}, {\it inflow traffic speed}, {\it wave temporal width}, {\it wave speed}, and {\it in-wave speed}, are systematically analyzed and discussed, by identifying their relative importance, characterizing their linear and nonlinear effects, and determining the direction of their influence on the JAD process (Table~\ref{tab:sensitivity_summary}). 
We further demonstrate that, within typical parameter ranges (Table~\ref{tab:parameter_ranges}), the JAD behavior is feasible in practice, as the entire process typically lasts only 30$\sim$160~s and spans approximately 1$\sim$13~km (Fig.~\ref{tab:sensitivity_summary}). 
Using a SUMO simulation as an illustrative example, we demonstrate and discuss how these parameters can be measured in practice with two stationary roadside traffic detectors. 
The results show that the proposed JAD strategy successfully suppresses the propagation of a stop-and-go wave, particularly without triggering secondary waves.
Although it is still a simulation-based validation, we argue that the results are convincing for the following two reasons. (i) As shown in Table~\ref{tab:Krauss}, stochasticity is explicitly incorporated into the modeling of driving behavior. (ii) As shown in Figure~\ref{fig:jad_trajectory}, the proposed strategy exhibits strong robustness, i.e., it can effectively suppress the propagation of traffic waves without requiring highly precise execution.

Among these parameters, determining an appropriate JAD speed that does not trigger traffic breakdown remains challenging in practice. 
This issue requires further advances in traffic-flow stability research, particularly empirical studies. 
Nevertheless, this limitation does not prevent the practical implementation of the proposed JAD strategy. 
In practice, a relatively higher JAD speed can be adopted: based on our experience and as demonstrated in this paper, speeds such as 50 km/h or even 70 km/h (as exhibited by real-world police-car swerving behavior shown in Fig.~\ref{fig:swerving}) are likely to be feasible.

Learning lessons from SPECIALIST \cite{VSL2008,VSL2010}, to ensure effectiveness and avoid unintended disturbances to the traffic system, it is essential to verify or predict, prior to dispatching a JAD vehicle, that the observed stop-and-go wave satisfies the required conditions. 
Specifically, the wave should be an isolated wave propagating downstream, rather than periodic oscillations associated with a fixed-location bottleneck. 
In short, a JAD vehicle should be dispatched only when a qualifying isolated stop-and-go wave is detected.

Overall, we believe that the present analysis substantially advances JAD studies from a purely theoretical concept toward practical application. 
In particular, we propose leveraging police-car swerving behavior to implement JAD, avoiding the common assumption in many previous studies that CAVs are required. 
While CAVs are often assumed in theoretical research nowadays, in reality, fully controllable vehicles that can participate in everyday traffic management are still rarely seen. 
With the proposed JAD strategy incorporating various practical considerations, the next step is real-world testing.

\section*{Acknowledgement}

We would like to thank Dr. Andreas Hegyi at TU Delft for sharing his valuable insights and practical experience with VSL and SPECIALIST, which helped inspire the ideas behind this, and Teng He for carefully reading the manuscript and identifying typos during the revision process.


\bibliographystyle{IEEEtran}
\bibliography{library}

\vspace{5cm}

\begin{IEEEbiography}[{\includegraphics[width=1in,height=1.25in,clip,keepaspectratio]{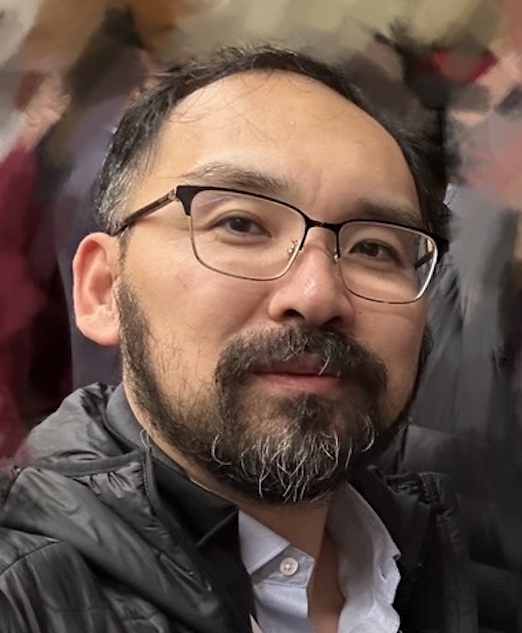}}] 
{Zhengbing He} (M'17-SM'20) received the Bachelor of Arts degree in English language and literature from Dalian University of Foreign Languages, China, in 2006, and the Ph.D. degree in systems engineering from Tianjin University, China, in 2011. 
He was a Post-Doctoral Researcher and an Assistant Professor with Beijing Jiaotong University, China. From 2018 to 2022, he was a Professor at Beijing University of Technology, China. 
From 2023 to 2025, he was a research scientist with Massachusetts Institute of Technology, USA. 
Presently, he is a Professor with University of Nottingham Ningbo China. 

\vspace{-2mm}

His research lies at the intersection of urban mobility, systems engineering, and artificial intelligence, spanning from traditional topics such as traffic flow operations and control, sustainability, and resilience to emerging areas including data-driven modeling, autonomous driving, and large language models. In particular, he is a pioneer and long-term contributor in AI-based transportation modeling and AV-empowered traffic congestion mitigation, and an early innovator in generative AI applications in transportation

\vspace{-2mm}

He has published more than 180 papers, including over 50 published exclusively in the prestigious journal series of Transportation Research and IEEE TRANSACTIONS (20+ TRC and 10+ TITS) and a Correspondence in Nature, with total citations exceeding 8,000 and H-index of over 40.
He was listed as the World’s Top 2\% Scientists, ranking 67th out of over 30,000 researchers in the field of Logistics and Transportation. He is the Editor-in-Chief of the Journal of Transportation Engineering and Information (Chinese). Meanwhile, he serves as a Senior Editor for IEEE TRANSACTIONS ON INTELLIGENT TRANSPORTATION SYSTEMS, an Associate Editor for IEEE TRANSACTIONS ON INTELLIGENT VEHICLES, a Deputy Editor-in-Chief of IET Intelligent Transport Systems, and an Editorial Advisory Board Member for Transportation Research Part C. His webpage is https://www.GoTrafficGo.com.
\end{IEEEbiography}

\end{document}